\documentclass[aps,pra,twocolumn,superscriptaddress,showpacs,showkeys,amsmath,amssymb]{revtex4}

\usepackage{amsfonts}
\usepackage{amssymb,amsmath}
\usepackage{mathrsfs}
\usepackage{latexsym}
\usepackage{amsmath}
\usepackage[cp1251]{inputenc}
\usepackage{graphicx}
\usepackage{dcolumn}
\usepackage{bm}
\usepackage{color}

\begin{document}

    \title{Impurity states in the one-dimensional Bose gas}

    \author{Volodymyr~Pastukhov\footnote{e-mail: volodyapastukhov@gmail.com}}
    \affiliation{Department for Theoretical Physics, Ivan Franko National University of Lviv,\\ 12 Drahomanov Street, Lviv-5, 79005, Ukraine}

    \date{\today}



    \begin{abstract}
   The detailed study of the low-energy spectrum for a mobile impurity in the one-dimensional bosonic environment is performed. Particularly we have considered only two analytically accessible limits, namely, the case of an impurity immersed in a dilute Bose gas, where one can use many-body perturbative techniques for low-dimensional bosonic systems, and the case of the Tonks–Girardeau (TG) gas, for which the usual fermionic diagrammatic expansion up to the second order is applied.
    \end{abstract}

    \maketitle

\section{Introduction}
\label{sec1}
\setcounter{equation}{0}
The behavior of impurity in the various mediums is a cornerstone problem for understanding of numerous phenomena in the condensed matter physics including Kondo effect, Anderson localization, etc. Recently, a great attention of theorists \cite{Novikov_Ovchinnikov_1,Novikov_Ovchinnikov_2,Rath,Li,Levinsen,Christensen,Grusdt_et_al,Vlietinck_et_al,Ardila_1,Ardila_2,Kain_Ling} has been paid to the analysis of a mobile impurity properties in the Bose gas. Such a renascence of the old problem well-studied in the context of a single $^3$He atom immersed in liquid $^4$He (see, for instance \cite{Krotscheck_et_al,Boronat,Galli,Panochko_et_al}, and references there) is stimulated by the success of the experimental techniques \cite{Hu,Jorgenzen} where the possibility to control a small amount of impurity particles strongly coupled to the bosonic bath is demonstrated.

A very interesting platform for the theoretical research in the Bose polaron problem is the case of one-dimensional environments \cite{Caldeira,Castro_Neto,Gangardt,Ovchinnikov,Matveev,Burovski,Dehkharghani,Petkovic}. It is well-known that due to highly non-trivial physics in the one spatial dimension \cite{Cazalilla_et_al,Imambekov_et_al} these systems possess unexpected behavior which very often obstructs their analysis. In some limiting cases, however, they admit analytical treatment \cite{Volosniev} or even existence of exact solutions \cite{McGuire_65,McGuire_66,Castella_Zotos,Gamayun_15,Gamayun_16}. There is also an experimental realization of the one-dimensional Bose polaron \cite{Catani_et_al}, where a minority of $^{41}$K atoms immersed in the $^{87}$Rb medium was observed during expansion and the prediction for the impurity effective mass within Feynman's framework was given. Essentially exact recent Monte Carlo simulations \cite{Parisi} revealed the impact of the considerable strong phonon-mediated interaction on the properties of a one-dimensional Bose polaron and to describe the system properly one needs to go beyond \cite{Grusdt} the Fr\"ohlich model in this case.

\section{Formulation}
We study the properties of a single impurity atom immersed in the Lieb-Liniger gas. It is assumed that the impurity interacts with bath particles via contact potential and by choosing appropriately a sign of the coupling constant we reproduce both repulsive and attractive Bose polarons. In order to take an advantage of many-body perturbation theory we consider the Bose-Fermi mixture consisting of a very dilute spinless (spin-polarized) Fermi gas immersed in the bosonic medium. The described model is characterized by the following Hamiltonian
\begin{eqnarray}\label{H}
	H=H_0+H_B+H_{int},
\end{eqnarray}
where $H_0$ describes ideal Fermi gas ($m_i$ is the mass of impurity particle)
\begin{eqnarray}\label{H_0}
	H_0=-\frac{\hbar^2 }{2m_i}\int^L_0 dx
	\,\psi^+(x)\partial^2_x\psi(x).
\end{eqnarray}
Here fermionic creation $\psi^+(x)$ and annihilation $\psi(x)$ field-operators which refer to the impurity states and satisfy usual anti-commutating relations
$\{\psi(x),\psi^+(x')\}=\delta(x-x')$, $\{\psi(x),\psi(x')\}=0$.
The second term in Eq.~(\ref{H}) is the Hamiltonian of Bose
particles of mass $m$ interacting with the $\delta$-like repulsive
potential
\begin{eqnarray}\label{H_B}
	H_B=-\frac{\hbar^2 }{2m}\int^L_0
	dx\,\phi^+(x)\partial^2_x\phi(x)\nonumber\\
	+\frac{g}{2}\int^L_0 dx\,(\phi^+(x))^2(\phi(x))^2,
\end{eqnarray}
where we have introduced field operators $\phi^+(x)$, $\phi(x)$ of Bose
type. Finally, the last term of $H$ takes into account the
interaction of Bose-Fermi subsystems
\begin{eqnarray}\label{H_int}
	H_{int}=\tilde{g}\int^L_0 dx\,\psi^+(x)\psi(x)\phi^+(x)\phi(x).
\end{eqnarray}
It is well-known that the formulated model (\ref{H}) can be exactly solved within the Bethe ansatz only when $m_i=m$, otherwise some approximate calculational schemes should be applied. But this equal-mass limit is a good benchmark for any perturbative approaches. In the following sections we will consider two opposite models of environments given by Hamiltonian (\ref{H_B}), namely, a dilute Bose gas (BEC) $g\rightarrow 0$, and a case of the TG limit $g\rightarrow \infty$. 

\subsection{Impurity in the dilute Bose gas}
For the low-dimensional systems $D\le 2$, where the condensate
does not exist at finite temperatures it is convenient to
introduce the phase-density representation
\cite{Mora_Castin,Cazalilla_et_al,Pastukhov_InfraredStr} for the bosonic operators:
$\phi(x)=e^{i\varphi(x)}\sqrt{n(x)}$,
$\phi^+(x)=\sqrt{n(x)}e^{-i\varphi(x)}$ with commutator
$[n(x),\varphi(x')]=i\delta(x-x')$ for the phase
$\varphi(x)$ and density $n(x)=\phi^+(x)\phi(x)$ fields. Imposing
periodic boundary conditions $n(x+L)=n(x)$, $\phi(x+L)=\phi(x)$
with large "volume'' $L$ and making use of the Fourier transform
$n(x)=n+\frac{1}{\sqrt{L}}\sum_{k\neq 0}e^{i kx}n_{k}$,
$\varphi(x)=\frac{1}{\sqrt{L}}\sum_{ k\neq 0}e^{-i kx}\varphi_{k}$,
where $n=N/L$ is the equilibrium density of Bose system;
substituting $\varphi(x)$, $n(x)$ in Eq.~(\ref{H_B}), and then
performing canonical transformation
$b_{k}=i\sqrt{n/\alpha_{k}}\,\varphi_{-
	k}+\frac{1}{2}\sqrt{\alpha_{k}/n}\,n_k$,
$b^+_k=-i\sqrt{n/\alpha_{k}}\,\varphi_{
	k}+\frac{1}{2}\sqrt{\alpha_{k}/n}\,n_{-k}$ (note that $[b_k,
b^+_q]=\delta_{k, q}$ and $[b_k, b_q]=0$) that diagonalizes
quadratic part of the Hamiltonian $H_B$ we finally obtain
\begin{eqnarray}\label{H_B_b}
	H_B=E_0+\sum_{k\neq 0}E_k b^+_kb_k+\Delta H_B,
\end{eqnarray}
\begin{eqnarray}
\Delta H_B=\frac{1}{3!\sqrt{N}}\sum_{k+q+s=0}D_{bbb}(k,q,s)b_k b_q b_s+\textrm{h.c.}\nonumber\\
+\frac{1}{2\sqrt{N}}\sum_{k,q\neq 0}D_{b^{\small
			+}bb}(k+q|k,q)b^+_{k+q} b_k
	b_q+\textrm{h.c.},
\end{eqnarray}
where $E_0$ and $E_k$ are the Bogoliubov ground-state energy and
quasiparticle spectrum, respectively. Introducing bosonic
free-particle dispersion $\varepsilon_k=\hbar^2k^2/2m$ one may
show that the above-mentioned requirement of diagonalization fixes
parameter $\alpha_k=E_k/\varepsilon_k$. It should be noted
that in $\Delta H_B$ the only relevant terms for our two-loop
calculations are presented. The functions
\begin{eqnarray}\label{D}
	\left.\begin{array}{c}
		D_{bbb}(k,q,s)\\
		D_{b^{\small +}bb}(s|k,q)
	\end{array}\right\}
	=\frac{\hbar^2}{4m\sqrt{\alpha_k\alpha_q\alpha_s}} \left[kq(\alpha_k\alpha_q+1)\right.\nonumber\\
	\left.+ks(\alpha_k\alpha_s\pm 1)+qs(\alpha_q\alpha_s\pm
		1)\right],
\end{eqnarray}
describe the simplest scattering processes of the elementary
excitations. In the same fashion we rewrite the third term of the
Hamiltonian
\begin{eqnarray}\label{H_int_b}
	&&H_{int}=n\tilde{g}\sum_p\psi^+_p\psi_p\nonumber\\
    &&+\frac{1}{\sqrt{L}}\sum_{p;
		k\neq 0}\tilde{g}\sqrt{n/\alpha_{k}}(b^+_k+b_{-
		k})\psi^+_p\psi_{p+k},
\end{eqnarray}
where operators $\psi^+_p$ and $\psi_p$ are the Fourier transform
of $\psi^+(x)$ and $\psi(x)$, respectively. Although we are going
to discuss the ground-state properties of the impurity atom
immersed in a Bose gas for the further analysis we adopt
the field-theoretical formulation at finite temperatures
\cite{Abrikosov}. The exact single-particle Green's function of
fermions in the four-momentum space is given by
\begin{eqnarray}
	G^{-1}_i(P)=i\nu_p-\xi_i(p)-\Sigma(P),
\end{eqnarray}
where $P=(\nu_p,p)$ ($\nu_p$ is the fermionic Matsubara
frequency); $\xi_i(p)=\hbar^2p^2/2m_i-\tilde{\mu}_i$, where the chemical potential
of the Fermi gas $\tilde{\mu}_i=\mu_i-n\tilde{g}$ shifted due to interaction with Bose subsystem ensures the particle number conservation. The exact self-energy of impurity is given by two skeleton diagrams depicted in
Fig.~1.
\begin{figure}[h!]
	\centerline{\includegraphics
		[width=0.40\textwidth,clip,angle=-0]{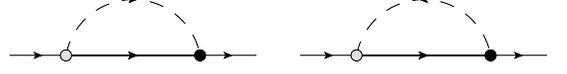}} \caption{Exact
		diagrammatic representation of self-energy $\Sigma(P)$ in the
		weakly-interacting Bose gas. Bold solid line represents the exact
		one-particle fermionic Green's function. Dashed line is the
		bosonic propagator in the Bogoliubov approximation. Dots stand for the
		zero-order (light) and the exact (black) vertices, respectively.}
\end{figure}
\begin{eqnarray}\label{Sigma_BEC}
&&	\Sigma(P)=\frac{-\tilde{g}}{L\beta}\sum_{K
	}\sqrt{\frac{n}{\alpha_k}}\,\Gamma_{b^{\small
		+}}(P-K,P)G_B(K)G_i(P-K)\nonumber\\
&&-\frac{\tilde{g}}{L\beta}\sum_{K
}\sqrt{\frac{n}{\alpha_k}}\,\Gamma_{b}(P+K,P)G_B(K)G_i(K+P),
\end{eqnarray}
where we have already took into account diluteness of the Bose
gas, i.e., neglected the self-energy corrections to the Green's
function $G_B(K)$ of Bogoliubov's quasiparticles.

This is formally an exact equation that determines the impurity
Green's function self-consistently. Technically this program for a
given approximation of the boson-fermion vertices
$\Gamma_{b^{\small +}}(P-K,P)$ and $\Gamma_{b}(P+K,P)$ is very
hard for practical realization, therefore, in the following we will
use perturbation theory. The appropriative expansion parameter is the coupling constant $\tilde{g}$ that characterizes the intensity of the two-particle Bose-Fermi
interaction which we accept to be small in our calculations. Following this ideology
one readily mentions that the correction of order $\tilde{g}^2$ to
the self-energy $\Sigma^{(1)}(P)$ is fully determined by the first
diagram in Fig.~1 with $\Gamma_{b^{\small +}}(P-K,P)\rightarrow
\tilde{g}\sqrt{n/\alpha_k}$ and $G_i(P)\rightarrow
1/[i\nu_p-\xi_i(p)]$. In this approximation the second diagram
provides the nonzero contribution only at finite temperatures. On
the two-loop level which particularly contains $\tilde{g}^3$- and
$\tilde{g}^4$-terms of the impurity self-energy the situation is
more complicated because now we have to take into account six
diagrams (see Fig.~2)
\begin{figure}[h!]
	\centerline{\includegraphics
		[width=0.4
		\textwidth,clip,angle=-0]{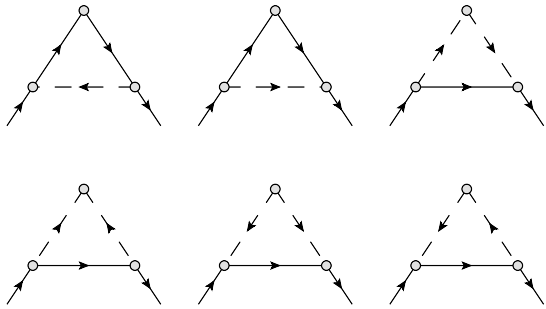}}
	\caption{One-loop diagrams contributing to the vertices $\Gamma_b(P+K,P)$, $\Gamma_{b^{\small +}}(P-K,P)$.}
\end{figure}
for each vertex $\Gamma_b(P+K,P)$,
$\Gamma_{b^{\small +}}(P-K,P)$ and also to use the impurity Green's function complicated with the first-order correction $G_i(P)=
1/[i\nu_p-\xi_i(p)]+\Sigma^{(1)}(P)/[i\nu_p-\xi_i(p)]^2$ in the first
diagram in Fig.~1. The details of these calculations as well as the explicit formula for
the self-energy up to the second order of a perturbation theory can
be found in Appendix A. Finally, it should be noted that the obtained in this section second-order formula for the self-energy can be applied to the Bose polaron problem in higher dimensions, for instance, in the three-dimensional case it reproduces results of Ref.~\cite{Christensen}.

\subsection{The Tonks–Girardeau limit}
Another interesting limit where the perturbative calculations may
be performed analytically is the case of the impurity immersed in
the Bose gas with infinite ($g\rightarrow \infty$) interparticle
repulsion. In this limit the operators $\phi^+(x)$, $\phi(x)$ by means of the Jordan-Wigner transformation can be mapped onto fermionic creation and annihilation field operators. Therefore in the following we have to consider the properties of the one-dimensional Fermi-Fermi mixture with unequal masses of two sorts of particles. The interaction is assumed to be switched on only between atoms of different species. The appropriate grand-canonical Hamiltonian
$H'=H-\sum_{p}\{\mu_i\psi^+_p\psi_p+\mu\phi^+_p\phi_p\}$ reads
\begin{eqnarray}\label{H_TG}
	H'=\sum_{p}\{\xi_i(p)\psi^+_p\psi_p+\xi_p\phi^+_p\phi_p\}\nonumber\\
	+\frac{1}{L}\sum_{p,q,k}\tilde{g}\,\psi^+_p\phi^+_q\phi_{q+k}\psi_{p-k},
\end{eqnarray}
where $\xi_p=\varepsilon_p-\mu$ with $\mu=\hbar^2p^2_0/2m$
($p_0=\pi n$) being the chemical potential of Bose gas in the
TG limit and we have to treat now $\phi^+_p$ and
$\phi_p$ as a Fermi creation and annihilation operators,
respectively. Then the impurity self-energy in the TG gas is (see
Fig.~3)
\begin{figure}[h!]
	\centerline{\includegraphics
		[width=0.20
		\textwidth,clip,angle=-0]{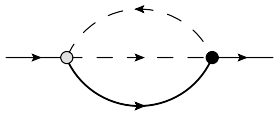}}
	\caption{The self-energy of the impurity atom immersed in the TG gas.
		Here black dot denotes exact vertex $T(Q-K;P+K|P;Q)$,
		while light dot stands for the Fourier transform $\tilde{g}$ of a bare interaction potential.}
\end{figure}
\begin{eqnarray}\label{Sigma_TG}
	\Sigma(P)=-\frac{\tilde{g}}{(L\beta)^2}\sum_{K,Q
	}\,T(Q-K;P+K|P;Q)\nonumber\\
	\times G_0(Q)G_0(Q-K)G_i(P+K).
\end{eqnarray}
Here again we incorporated the Hartree term to the redefinition of
the impurity binding energy $\mu_i\rightarrow \tilde{\mu}_i$ and
introduced notation for the one-particle Green's function
$G_0(Q)=1/[i\nu_q-\xi_q]$ of a bosonic medium in the infinite-$g$
limit. Keeping in mind perturbative consideration in terms of the
impurity-boson coupling parameter we obtain the self-energy in the simplest
approximation by replacing the vertex $T(Q-K;P+K|P;Q)$ with
$\tilde{g}$. The second-order calculation requires both vertex
corrections presented in Fig.~4 and the one-loop self-energy
insertion in the internal impurity propagator of a diagram in
Fig.~3 (for more details see Appendix B).
\begin{figure}[h!]
	\centerline{\includegraphics
		[width=0.35
		\textwidth,clip,angle=-0]{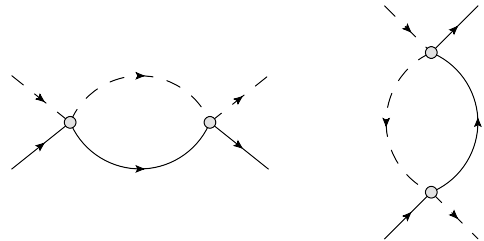}}
	\caption{One-loop diagrams contributing to the two-particle vertex $T(P;Q|Q+K;P-K)$.}
\end{figure}

In general, the impurity spectrum can be found from the poles of
retarded Green's function. Particularly for the real part of the
spectrum one obtains
\begin{eqnarray}\label{xi*}
	\xi^*_i(p)=\xi_i(p)+\Sigma_R(\xi^*_i(p),p),
\end{eqnarray}
where $\Sigma_R(\nu,p)=\Re \Sigma(P)|_{i\nu_p\rightarrow \nu+i0}$
is the real part of an analytically-continued in the upper complex
half-plane self-energy. Up to the second order of the perturbation
theory Eq.~(\ref{xi*}) reads
\begin{eqnarray}\label{xi*_2}
	\xi^*_i(p)=\xi_i(p)+\Sigma^{(1)}_R(\xi_i(p),p)+\Sigma^{(2)}_R(\xi_i(p),p)\nonumber\\
	+\frac{1}{2}\frac{\partial}{\partial \xi_i(p)}\left[\Sigma^{(1)}_R(\xi_i(p),p)\right]^2,
\end{eqnarray}
where $\Sigma^{(1)}_R(\nu,p)$ and $\Sigma^{(2)}_R(\nu,p)$ are real parts of the one-
and two-loop corrections to the self-energy, respectively. Absence
of the Fermi surface for the impurity atom guarantees that its spectrum is
gapless, i.e.,  $\xi^*_i(p\rightarrow 0)\rightarrow 0$. In the
long-wavelength limit it is characterized by the effective mass
only, and by expanding the right-hand side of the above equation
we are in position to calculate both the impurity biding energy
\begin{eqnarray}
	\mu_i=n\tilde{g}+\mu^{(1)}_i+\mu^{(2)}_i+\ldots,
\end{eqnarray}
and the inverse effective mass 
\begin{eqnarray}
	m_i/m^*_i=1+\Delta^{(1)}+\Delta^{(2)}+\ldots,
\end{eqnarray}
where the superscript denotes the order of perturbation theory.

\section{Results}
\subsection{One-loop calculations}
The general low-energy structure of the impurity Green's function
is visible even in the simplest approximation. Therefore it is
worthwhile to discuss the first-order result in more details
furthermore that these calculations in the small-$g$ limit can be
performed analytically. In particular, for the first correction to
the impurity binding energy, which is determined only by $\Sigma^{(1)}_R(-\tilde{\mu}_i,0)$
we obtained $\mu^{(1)}_i/(n\tilde{g})=\alpha\epsilon^{(1)}(\gamma)$,
where function $\epsilon^{(1)}(\gamma)$ of the mass ratio
$\gamma=m/m_i$ in the case of a weakly-interacting Bose gas
\begin{eqnarray}
\epsilon^{(1)}_{BEC}(\gamma)=-\frac{1}{\sqrt{\gamma^2-1}}\ln\left|\frac{\gamma+\sqrt{\gamma^2-1}}
{\gamma-\sqrt{\gamma^2-1}}\right|,
\end{eqnarray}
and in the TG limit 
\begin{eqnarray}
\epsilon^{(1)}_{TG}(\gamma)=-\int^{1}_0\frac{d q}{q}\ln\left|\frac{(1+q)^2+\gamma(1-q^2)}
{(1-q)^2+\gamma(1-q^2)}\right|,
\end{eqnarray}
is presented in Fig.~5.
\begin{figure}[h!]
	\centerline{\includegraphics
		[width=0.4
		\textwidth,clip,angle=-0]{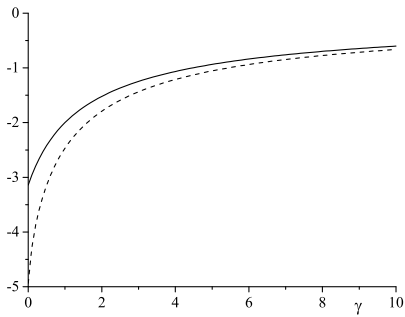}}
	\caption{Dimensionless correction $\epsilon^{(1)}(\gamma)$ to the impurity binding
		energy as a function of mass ratio $\gamma=m/m_i$ in the BEC (solid) and TG (dashed) limits.}
\end{figure}
The dimensionless coupling constant $\alpha=\tilde{g}/(2\pi\hbar
c)$ (with $c$ being the sound velocity in both cases, i.e,
$c=\sqrt{ng/m}$ and $c=\hbar p_0/m$ in BEC and TG limits,
respectively) is the expansion parameter which controls the limits
of applicability of our perturbative results. At finite momenta
the self-energy $\Sigma^{(1)}_R(\xi_i(p),p)$ in the BEC side is
non-monotonic function of the wave-vector with logarithmic
divergence at $p=m_ic/\hbar$, i.e., when the velocity of impurity
reaches the value of the velocity of sound propagation in the
bosonic system. Qualitatively the same behavior of the impurity
self-energy is observed in the TG gas. Furthermore, from the exact
solution of Lieb and Liniger model \cite{Lieb_Liniger} we learn
that the spectrum of system \cite{Lieb} contains two phonon-like
branches in the long-length limit for any finite value of coupling
constant $g$. Therefore these divergences always appear indicating
a non-perturbative nature of the impurity self-energy in the
momentum region close to $\hbar p=m_ic$. On the other hand it is
well-known that the impurity moving with supersonic velocity
starts to dissipate its energy by producing elementary excitations in the bosonic bath. In the one spatial dimension this dissipation is so intensive that the
imaginary part of the self-energy
$\Sigma^{(1)}_I(\xi_i(p),p)/(n\tilde{g})=-\pi \alpha/\gamma$,
($p=m_ic/\hbar+0$) is of order magnitude to the real one, in what
follow we cannot neglect the damping and use 
Eq.~(\ref{xi*_2}) to determine the impurity spectrum at this
point. But in the long-wavelength limit $p\ll m_ic/\hbar$ the
damping is absent so the perturbative impurity spectrum is
well-defined. The one-loop contribution to the effective mass is
given by $\Delta^{(1)}_{TG}=-4\alpha^2$ in the TG limit and by
\begin{eqnarray}
	&&\Delta^{(1)}_{BEC}=-\frac{\tilde{g}}{g}\frac{\alpha\gamma}{\gamma^2-1}\nonumber\\
	&&\times\left[2\gamma-
	\frac{1}{\sqrt{\gamma^2-1}}\ln\left|\frac{\gamma+\sqrt{\gamma^2-1}}
	{\gamma-\sqrt{\gamma^2-1}}\right| \right],
\end{eqnarray}
in the dilute one-dimensional Bose gas.

No less interesting is the behavior of the impurity wave-function
renormalization (quasiparticle residue)
$Z^{-1}_{i}(p)=1-\partial\Sigma_R(\xi_i(p),p)/\partial\xi_i(p)$.
It is easy to show by the direct calculations that the above
derivative is logarithmically divergent for {\it any} $p$ both in
the BEC and TG limits. Particularly it means that the series expansion of the inverse retarded Green's function, calculated in the first-order of perturbation theory, reads ($\nu\rightarrow \xi^{*}_i(p)$)
\begin{eqnarray}\label{G_pert}
	&&[G_i^{ret}(\nu,p)]^{-1}=\nu-\xi_i(p)-\Sigma^{(1)}_R(\nu,p)\rightarrow\nonumber\\
	&&\nu-\xi^{*}_i(p)-\frac{\partial\Sigma^{(1)}_R(\xi^{*}_i(p),p)}{\partial\xi^{*}_i(p)}[\nu-\xi^{*}_i(p)]+\ldots\rightarrow\nonumber\\
	&&[\nu-\xi^{*}_i(p)]\left\{1-\eta^{(1)}(p)\ln[\nu-\xi^{*}_i(p)]+\ldots \right\},
\end{eqnarray}
where dots stand for the finite terms. Being independent of the wave-vector these divergences suggest the {\it exact} Green's function to have a branch-point singularity ($\eta(0)=\eta$)
\begin{eqnarray}\label{G_as}
	G_i^{ret}(\nu,p\rightarrow 0)\big{|}_{\nu\rightarrow
		\xi^{*}_i(p)}\propto \frac{1}{[\nu-\xi^{*}_i(p)]^{1-\eta}}.
\end{eqnarray}
This statement is supported by the explicit calculation of the exponent
$\eta$ since in both analytically available cases we obtained the same value
\begin{eqnarray}\label{eta_1}
\eta^{(1)}_{BEC}=\eta^{(1)}_{TG}=n\tilde{g}^2/(2\pi\hbar mc^3),
\end{eqnarray}
on the one-loop level. Looking ahead it should be noted that this power-law behavior of the impurity Green's function is perfectly confirmed by the second-order perturbation theory calculations.

\subsection{Two-loop results}
The numerical calculations up to the second-order of perturbation theory requires more computational efforts. Particularly the expansion for binding energy correction in the BEC limit reads $\mu^{(2)}_i/n\tilde{g}=\frac{\tilde{g}}{g}\alpha^2\epsilon^{(2,1)}_{BEC}(\gamma)+\alpha^2\epsilon^{(2,2)}_{BEC}(\gamma)$. In the TG case the above expansion contains single term $\mu^{(2)}_i/(n\tilde{g})=\alpha^2\epsilon^{(2)}_{TG}(\gamma)$. For comparison in Fig.~6 we built all three curves. It is seen that function $\epsilon^{(2,2)}_{BEC}(\gamma)$ is almost two order magnitude larger than $\epsilon^{(2,2)}_{BEC}(\gamma)$ which particularly means that even in a weakly-interacting Bose gas the quasiparticle-mediated impurity potential is not negligible.
\begin{figure}[h!]
	\centerline{\includegraphics
		[width=0.4
		\textwidth,clip,angle=-0]{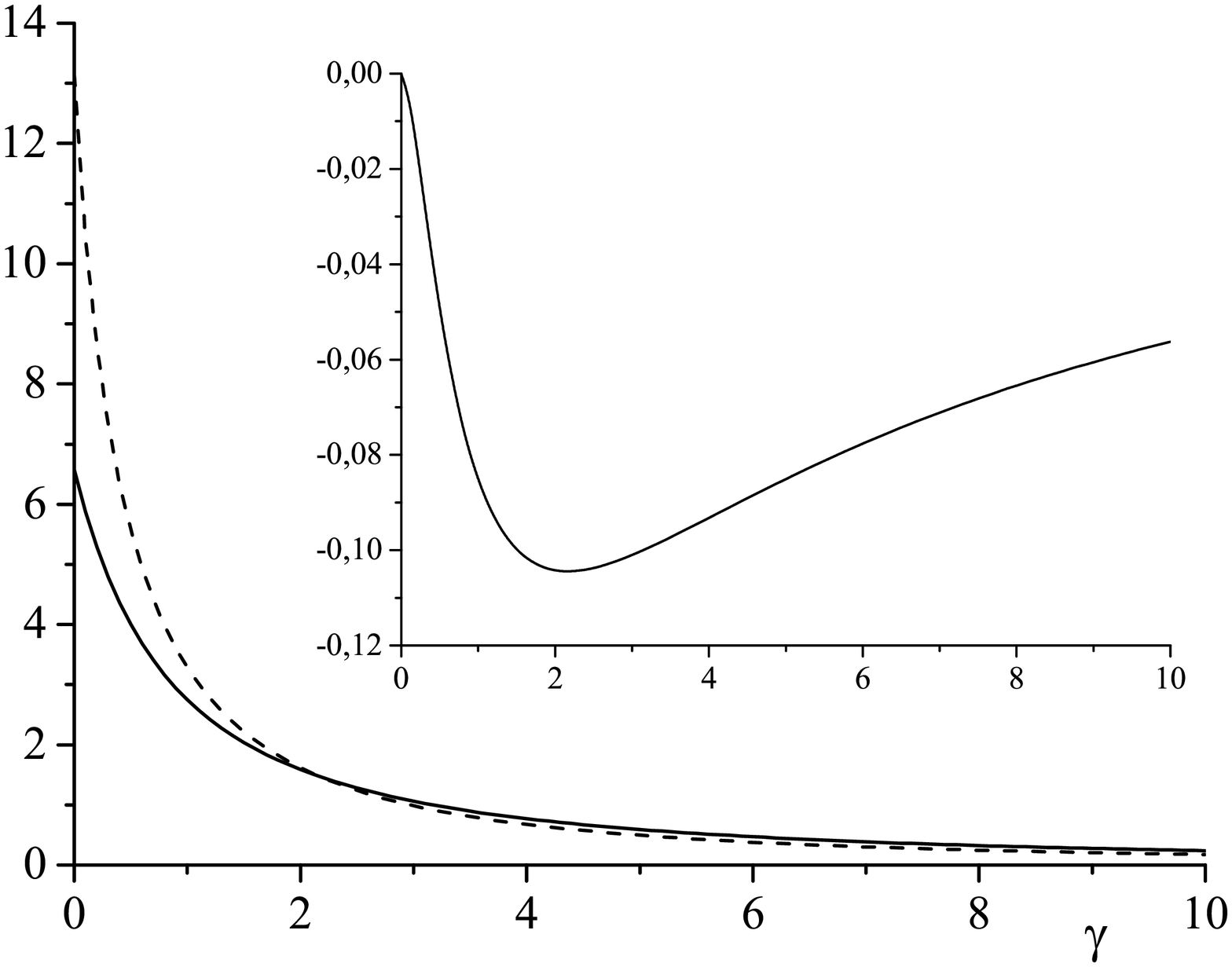}}
	\caption{The second-order binding energy corrections $\epsilon^{(2,2)}_{BEC}(\gamma)$ (solid) and $\epsilon^{(2)}_{TG}(\gamma)$ (dashed). Inset shows function $\epsilon^{(2,1)}_{BEC}(\gamma)$.}
\end{figure}
In the TG limit when $\gamma=1$ our results for the impurity binding energy $\mu_i|_{\gamma=1}=n\tilde{g}\left(1-\frac{\pi^2}{4}\alpha+\frac{\pi^2}{3}\alpha^2 +\ldots\right)$ exactly reproduces the first three terms of an analytical formula \cite{McGuire_65} obtained within Bethe ansatz wave-function. The similar expansions was derived for the second-order corrections to the particle effective mass in the BEC $\Delta^{(2)}_{BEC}=(\tilde{g}/g)^2\alpha^2\delta^{(2,1)}_{BEC}(\gamma)+(\tilde{g}/g)\alpha^2\delta^{(2,2)}_{BEC}(\gamma)$ and TG $\Delta^{(2)}_{TG}=\alpha^3\delta^{(2)}_{TG}(\gamma)$ limits, respectively (functions $\delta^{(2,1)}_{BEC}(\gamma)$, $\delta^{(2,2)}_{BEC}(\gamma)$ and $\delta^{(2)}_{TG}(\gamma)$ are plotted on Fig.~7 and Fig.~8).
\begin{figure}[h!]
	\includegraphics[scale=0.25]{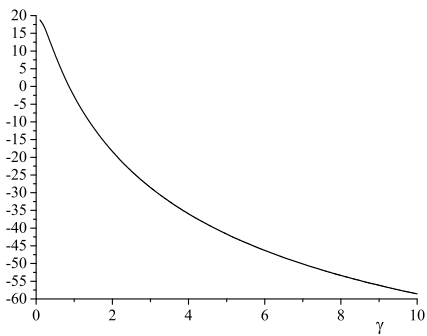}
	\caption{Function $\delta^{(2)}_{TG}(\gamma)$ determining the two-loop result for an effective mass in the TG limit.}
\end{figure}
\begin{figure}[h!]
	\includegraphics[scale=0.25]{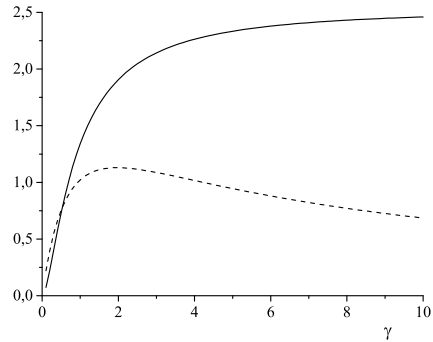}
	\caption{Dimensional functions $\delta^{(2,1)}_{BEC}(\gamma)$ (solid) and $\delta^{(2,2)}_{BEC}(\gamma)$ (dashed).}
\end{figure}
It is easy to verify that the numerically calculated TG effective mass in the integrable limit perfectly coincide with the exact expansion $m_i/m^*_i|_{\gamma=1}=1-4\alpha^2+4(\pi^2/3-4)\alpha^3+\ldots$. The figure~7 reveals the strong dependence of function $\delta^{(2)}_{TG}(\gamma)$ on the mass ratio parameter $\gamma$. This signals the break down of an ordinary perturbation theory at large mass imbalance in the TG limit and in order to resolve this problem one needs to take into account infinite series of diagrams (ladder summation in the particle-particle or particle-hole \cite{Gamayun} channels).

Our second-order perturbative calculations of the self-energy allow to obtain the above presented exponent $\eta$ on the two-loop level. In the same manner as it was done before (see Eq.~(\ref{G_pert})) by the explicit series expansion of the retarded impurity Green's function in the vicinity of a singular point $\nu\rightarrow \xi^{*}_i(p)$ we have
\begin{eqnarray*}
	[G_i^{ret}(\nu,p)]^{-1}=[\nu-\xi^{*}_i(p)]\left\{1-[\eta^{(1)}+\eta^{(2)}]\ln[\nu-\xi^{*}_i(p)]
	\right.\nonumber\\
	\left.+\frac{1}{2}[\eta^{(1)}]^2\ln^2[\nu-\xi^{*}_i(p)]\pm\ldots\right\},
\end{eqnarray*}
where $\eta^{(1)}$ was already given by Eq.~(\ref{eta_1}) and value of the second correction $\eta^{(2)}$ depends strongly on the properties of bosonic environment. The presence of the $\ln^2$-divergences with a proper factor $\frac{1}{2}[\eta^{(1)}]^2$ proves our original suggestion (\ref{G_as}). Combining the first- and second-order results we obtain in the BEC limit 
\begin{eqnarray*}
	&&\eta^{(1)}_{BEC}+\eta^{(2)}_{BEC}=\frac{n\tilde{g}^2}{2\pi\hbar mc^3}\nonumber\\
	&&\times\left[1- \frac{\alpha}{2\sqrt{\gamma^2-1}}\ln\left|\frac{\gamma+\sqrt{\gamma^2-1}}
	{\gamma-\sqrt{\gamma^2-1}}\right| \right]^2.
\end{eqnarray*}
The TG limit demonstrates an unexpected behavior $\eta^{(2)}_{TG}=0$ instead.
Such a dependence of the second-order correction $\eta^{(2)}$ led us to conclusion that the exact value of an exponent responsible for the non-analytic behavior of the impurity Green's function is given by
$\eta=n\left(\partial \mu_i/\partial n\right)^2/(2\pi\hbar mc^3)$. Indeed, it is easy to verify that $\mu^{(1)}_i$ that determines $\eta^{(2)}$ does not depend on density of the medium in the TG limit and that a derivative $\partial \mu^{(1)}_i/\partial (n\tilde{g})$ in the BEC side is equal to the expression in square brackets of Eq.~(\ref{eta_2}). Of course, it is too optimistic to write down the whole result only with the second-order perturbative calculations in hand but exactly the same formula for $\eta$ as well as a singular behavior (\ref{G_as}) of the impurity propagator can be proven by using a technique similar to that of Refs.~\cite{Pastukhov_15,Pastukhov_16}.

\section{Conclusions}
In summary, by applying perturbation theory up to the second order we have revealed the detailed low-energy structure of the spectrum (binding energy and effective mass) for a mobile impurity immersed in the one-dimensional bosonic environment. Considering our system as a Fermi-Bose mixture with the vanishingly small fermionic density we found that the interaction with bosonic medium crucially changes the single-particle impurity Green's function providing the latter to exhibit branch-point singularity. Using our second-order perturbative results we have proposed the general formula for the non-universal exponent determining this behavior. It is also demonstrated that the induced interaction especially in the case of a large mass imbalance has a profound effect on the behavior of a single impurity atom in the one-dimensional Bose gas.

\begin{center}
	{\bf Acknowledgements}
\end{center}

Stimulating discussions with Prof.~I.~Vakarchuk and Dr.~A.~Rovenchak are gratefully acknowledged. This work was partly supported by Project FF-30F (No.~0116U001539) from the Ministry of Education and Science of Ukraine.

\section{Appendices}
\subsection{BEC limit}
The calculations of the one-loop diagrams (see Fig.~1) in the BEC side at zero temperature yield
\begin{eqnarray}
	\Sigma^{(1)}(P)=-\frac{1}{L}\sum_{k\neq 0}\frac{n\tilde{g}^2}{\alpha_k}\frac{1}{E_k+\xi_i(k+p)-i\nu_p}.
\end{eqnarray}
The second-order result is more cumbersome for evaluation nevertheless still tractable. While calculating the one-loop correction to vertices we find that only four diagrams in Fig.~2 are non-zero at $T=0$. Furthermore, by substituting these eight terms in Eq.~(\ref{Sigma_BEC}) one concludes that only five contribute to the self-energy with the result:
\begin{widetext}
\begin{eqnarray}
	\Sigma^{(2)}(P)=-\frac{1}{2L^2}\sum_{k,s\neq 0}\frac{n^2\tilde{g}^4}{\alpha_k\alpha_s}\frac{1}{E_s+E_k+\xi_i(s+k+p)-i\nu_p}\left[\frac{1}{E_k+\xi_i(k+p)-i\nu_p}+\frac{1}{E_s+\xi_i(s+p)-i\nu_p}\right]^2\nonumber\\
	+\frac{1}{L^2}\sum_{k,s \neq 0}\frac{n\tilde{g}^3}{\alpha_k\alpha_s\alpha_{k+s}}\frac{1}{E_s+\xi_i(s-p)-i\nu_p}\frac{1}{E_k+\xi_i(k+p)-i\nu_p}\nonumber\\
	-\frac{1}{2L^2}\sum_{k,s \neq 0}\frac{n\tilde{g}^3}{\alpha_k\alpha_s\alpha_{k+s}}\frac{1}{E_k+E_s+\xi_i(k+s+p)-i\nu_p}\left[\frac{D_{+}(k,s)}{E_{k+s}+\xi_i(s+k+p)-i\nu_p}-\frac{D_{-}(k,s)}{E_k+E_s+E_{k+s}}\right]\nonumber\\
	\times \left[\frac{1}{E_k+\xi_i(k+p)-i\nu_p}+\frac{1}{E_s+\xi_i(s+p)-i\nu_p}\right],
\end{eqnarray}
where the symmetric functions $D_{\pm}(k,s)$ read
\begin{eqnarray*}
	D_{\pm}(k,s)=\frac{\hbar^2}{2m}\left[k(k+s)(\alpha_k-1)(\alpha_{k+s}\pm1)+s(s+k)(\alpha_s-1)(\alpha_{k+s}\pm1)\pm ks(\alpha_k-1)(\alpha_s-1)\right].
\end{eqnarray*} 
\end{widetext}
\subsection{TG gas}
The self-energy calculations in the TG limit is much simpler. For instance, on the one-loop level we obtained
\begin{eqnarray}
	\Sigma^{(1)}(P)=\frac{1}{L}\sum_{q}\tilde{g}^2(1-n_q)\Pi_q(P)\nonumber\\
	=-\frac{1}{L}\sum_{q}\tilde{g}^2 n_qt_q(P),
\end{eqnarray}
where $n_q=\theta(p_0-|q|)$ is a unit step-function. The impurity-boson particle-hole diagram reads
\begin{eqnarray}
	\Pi_q(P)=\frac{1}{L}\sum_{k}\frac{n_k}{i\nu_p-\xi_q+\xi_k-\xi_i(k-q+p)},
\end{eqnarray}
and notation for particle-particle bubble
\begin{eqnarray}
	t_q(P)=\frac{1}{L}\sum_{k}\frac{1-n_k}{\xi_k+\xi_i(k+q+p)-\xi_q-i\nu_p},
\end{eqnarray}
is used. Taking into account the vertex corrections (see Fig.~4) the calculations in the next order of perturbation theory give
\begin{eqnarray}
	\Sigma^{(2)}(P)=\frac{1}{L}\sum_{q}\tilde{g}^3n_qt^2_q(P)\nonumber\\-\frac{1}{L}\sum_{q}\tilde{g}^3(1-n_q)\Pi^2_q(P).
\end{eqnarray}
Both $\Pi_q(P)$ and $t_q(P)$ on the ``mass-shell'' $i\nu_p\rightarrow \xi_i(p)$ can be written in terms of elementary functions. It is easy to argue that the contribution of the self-energy insertion to the impurity spectrum is of order $\tilde{g}^4$ and therefore is neglected in the present study.

\end{document}